\documentclass[12pt]{article}
\usepackage{amssymb}
\makeatletter
\def\numberbysection{\@addtoreset{equation}{section}
 	\def\theequation{\thesection.\arabic{equation}}}
\makeatother

\numberbysection

\newcommand{\be}{\begin{eqnarray}}
\newcommand{\ee}{\end{eqnarray}}
\newcommand{\non}{\nonumber}
\newcommand{\tr}{\mathop{\rm tr}\nolimits}

\newcommand{\kk}{\kappa}
\newcommand{\id}{\mathbb{I}}

\begin{document}

\begin{titlepage}
\strut\hfill UMTG--232
\vspace{.5in}
\begin{center}

\LARGE Solving the open XXZ spin chain with 
nondiagonal boundary terms at roots of unity\\[1.0in]
\large Rafael I. Nepomechie\\[0.8in]
\large Physics Department, P.O. Box 248046, University of Miami\\[0.2in]  
\large Coral Gables, FL 33124 USA\\

\end{center}

\vspace{.5in}

\begin{abstract}
\end{abstract}
We consider the open XXZ quantum spin chain with nondiagonal boundary
terms.  For bulk anisotropy value $\eta = {i \pi\over p+1}$, $p= 1
\,, 2 \,, \ldots $, we propose an exact $(p+1)$-order functional
relation for the transfer matrix, which implies Bethe-Ansatz-like
equations for the corresponding eigenvalues.  The key observation is
that the fused spin-${p+1\over 2}$ transfer matrix can be expressed in
terms of a lower-spin transfer matrix, resulting in the truncation of
the fusion hierarchy.
\end{titlepage}

\setcounter{footnote}{0}

\section{Introduction}\label{sec:intro}

A long outstanding problem has been to solve the open XXZ quantum spin
chain with nondiagonal boundary terms, defined by the Hamiltonian
\cite{Sk, dVGR, GZ}
\be
{\cal H }&=& {1\over 2}\Big\{ \sum_{n=1}^{N-1}\left( 
\sigma_{n}^{x}\sigma_{n+1}^{x}+\sigma_{n}^{y}\sigma_{n+1}^{y}
+\cosh \eta\ \sigma_{n}^{z}\sigma_{n+1}^{z}\right)\non \\
&+&\sinh \eta \Big( \coth \xi_{-} \sigma_{1}^{z}
+ {2 \kk_{-}\over \sinh \xi_{-}}\sigma_{1}^{x} 
- \coth \xi_{+} \sigma_{N}^{z}
- {2 \kk_{+}\over \sinh \xi_{+}}\sigma_{N}^{x} \Big) \Big\} \,,
\label{Hamiltonian}
\ee
where $\sigma^{x} \,, \sigma^{y} \,, \sigma^{z}$ are the standard
Pauli matrices, $\eta$ is the bulk anisotropy parameter, $\xi_{\pm} \,,
\kk_{\pm}$ are arbitrary boundary parameters, and $N$ is the number of 
spins.  Solving this problem (e.g., determining the Bethe Ansatz
equations) is a crucial step in formulating the thermodynamics of the
spin chain and of the boundary sine-Gordon model.  Moreover, this
problem has important applications in condensed matter physics and
statistical mechanics.

A fundamental difficulty is that, in contrast to the special case of
diagonal boundary terms (i.e., $\kk_{\pm}=0$) considered in
\cite{ABBBQ, Sk}, a simple pseudovacuum state does not exist.  Hence,
most of the techniques which have been developed to solve integrable
models cannot be applied.  Moreover, it is not yet clear how to
implement the few techniques (such as Baxter's $T-Q$ approach
\cite{Ba1} or the generalized algebraic Bethe Ansatz \cite{Ba1, FT}) which 
do not rely on a pseudovacuum state.

We report here some progress on this problem.  Namely, for bulk
anisotropy value
\be
\eta = {i \pi\over p+1}\,, \qquad p= 1 \,, 2 \,, \ldots \,,
\label{etavalues}
\ee
(and hence $q \equiv e^{\eta}$ is a root of unity, satisfying
$q^{p+1}=-1$), we propose an exact $(p+1)$-order functional relation
for the fundamental transfer matrix, which implies Bethe-Ansatz-like
equations for the corresponding eigenvalues.  The key observation is
that the fused transfer matrices $t^{(j)}(u)$, which are constructed with
a spin-$j$ auxiliary space, satisfy the identity \footnote{This
is distinct from the observation due to Belavin {\it et al.} \cite{BS,
BGF} that, for the special case of quantum-group symmetry (i.e.,
$\kk_{\pm}=0$, $\xi_{\pm}\rightarrow \infty$), the fused transfer matrix
$t^{({p\over 2})}(u)$ vanishes after quantum group reduction.}
\be
t^{({p+1\over 2})}(u) = \alpha(u) \left[ t^{({p-1\over 2})}(u+ \eta) + 
\beta(u) \id \right]
\,, \label{truncation}
\ee
where $t^{(0)} = \id$ (identity matrix), and $ \alpha(u)\,, \beta(u)$
are scalar functions.  That is, the spin-${p+1\over 2}$ transfer
matrix can be expressed in terms of a lower-spin transfer matrix,
resulting in the truncation of the fusion hierarchy.  We have verified
this result explicitly for $p = 1 \,, 2\,, 3$, and we conjecture that
it is true for all positive integer values of $p$.  The simplest case
$p=1$, which corresponds to the XX chain, has recently been analyzed
in \cite{NeXX}.  Similar higher-order functional relations have been
obtained for the closed (periodic boundary conditions) 8-vertex model
by Baxter \cite{Ba2} using a different method.

The outline of this article is as follows.  In Section
\ref{sec:transfer} we review the construction of the fundamental
($j={1\over 2}$) transfer matrix which contains the
Hamiltonian (\ref{Hamiltonian}).  In Section \ref{sec:fusion} we
briefly review the so-called fusion procedure \cite{Ka}-\cite{IOZ} and the
construction of the higher-spin transfer matrices, which obey an
infinite fusion hierarchy.  In Section \ref{sec:truncation}, we obtain
the relation (\ref{truncation}), to which we refer as the ``truncation
identity,'' since it serves to truncate the fusion hierarchy.  In
Section \ref{sec:functional} we present the exact functional relations
which are obeyed by the fundamental transfer matrix.  The corresponding
sets of Bethe Ansatz equations are given in Section
\ref{sec:eigenvalues}.  We conclude with a brief discussion of our
results in Section \ref{sec:discuss}.

\section{Fundamental transfer matrix}\label{sec:transfer}

We recall \cite{Sk} that the transfer matrix for an open chain is 
made from two basic building blocks, called $R$ (bulk) and $K$ 
(boundary) matrices.

An $R$ matrix is a solution of the Yang-Baxter equation 
\be
R_{12}(u-v)\ R_{13}(u)\ R_{23}(v)
= R_{23}(v)\ R_{13}(u)\ R_{12}(u-v) \,.
\label{YB}
\ee 
(See, e.g., \cite{KS, KBI, NePrimer}.) For the XXZ spin chain, 
the $R$ matrix is the $4 \times 4$ matrix
\be
R(u) = \left( \begin{array}{cccc}
	\sinh  (u + \eta) &0            &0           &0            \\
        0                 &\sinh  u     &\sinh \eta  &0            \\
	0                 &\sinh \eta   &\sinh  u    &0            \\
	0                 &0            &0           &\sinh  (u + \eta)
\end{array} \right) \,,
\label{bulkRmatrix}
\ee 
where $\eta$ is the anisotropy parameter.  This $R$ matrix has the
symmetry properties
\be
R_{12}(u) = {\cal P}_{12} R_{12}(u) {\cal P}_{12} 
= R_{12}(u)^{t_{1} t_{2}} \,,
\ee
where ${\cal P}_{12}$ is the permutation matrix and $t$ denotes 
transpose.  Moreover, it satisfies the unitarity relation
\be
R_{12}(u)\ R_{12}(-u) = \zeta(u) \id \,, \qquad 
\zeta(u)=-\sinh(u+\eta) \sinh(u-\eta) \,,
\label{unitarity}
\ee
and the crossing relation 
\be
R_{12}(u) = V_{1} R_{12}(-u-\eta)^{t_{2}} V_{1} \,, 
\quad V =  -i\sigma^{y} \,.
\label{crossing}
\ee 
Finally, it has the periodicity property
\be
R_{12}(u+i \pi)= - \sigma^{z}_{2} R_{12}(u) \sigma^{z}_{2} =
- \sigma^{z}_{1} R_{12}(u) \sigma^{z}_{1} \,.
\label{periodicityR}
\ee 

The matrix $K^{-}(u)$ is a solution of the boundary Yang-Baxter equation 
\cite{Ch}
\be
R_{12}(u-v)\ K^{-}_{1}(u)\ R_{21}(u+v)\ K^{-}_{2}(v)
= K^{-}_{2}(v)\ R_{12}(u+v)\ K^{-}_{1}(u)\ R_{21}(u-v) \,.
\label{boundaryYB}
\ee 
We consider here the following $2 \times 2$ matrix \cite{dVGR, GZ}
\be
K^{-}(u) = \left( \begin{array}{cc}
\sinh(\xi_{-} + u)   & \kk_{-} \sinh  2u \\
\kk_{-} \sinh  2u     & \sinh(\xi_{-} - u) 
\end{array} \right) \,,
\label{Kminusmatrix}
\ee 
which evidently depends on two boundary parameters $\xi_{-} \,, 
\kk_{-}$.  We set the matrix $K^{+}(u)$ to be $K^{-}(-u-\eta)$ with 
$(\xi_{-} \,, \kk_{-})$ replaced by $(\xi_{+} \,, \kk_{+})$; i.e.,
\be
K^{+}(u) = \left( \begin{array}{cc}
-\sinh(u + \eta - \xi_{+})   &  -\kk_{+} \sinh  (2u +2\eta) \\
-\kk_{+} \sinh  (2u +2\eta)    & \sinh(u + \eta + \xi_{+})  
\end{array} \right) \,.
\label{Kplusmatrix}
\ee 
The $K$ matrices have the periodicity property
\be
K^{\mp}(u + i \pi) = - \sigma^{z} K^{\mp}(u)  \sigma^{z} \,.
\label{periodicityK}
\ee 

The fundamental transfer matrix $t(u)$ for an open chain of $N$ spins
is given by \cite{Sk}
\be
t(u) = \tr_{0} K^{+}_{0}(u)\  
T_{0}(u)\  K^{-}_{0}(u)\ \hat T_{0}(u)\,,
\label{transfer}
\ee
where $\tr_{0}$ denotes trace over the ``auxiliary space'' 0,
and $T_{0}(u)$, $\hat T_{0}(\lambda)$ are so-called monodromy 
matrices \footnote{As is customary, we usually suppress the 
``quantum-space'' subscripts $1 \,, \ldots \,, N$.}
\be
T_{0}(u) = R_{0N}(u) \cdots  R_{01}(u) \,,  \qquad 
\hat T_{0}(u) = R_{10}(u) \cdots  R_{N0}(u) \,. 
\label{monodromy}
\ee
Indeed, Sklyanin has shown that $t(u)$ constitutes a one-parameter 
commutative family of matrices
\be
\left[ t(u)\,, t(v) \right] = 0  \,.
\label{commutativity}
\ee 
The Hamiltonian (\ref{Hamiltonian}) is related to the first derivative
of the transfer matrix
\be
{\cal H} = {t'(0)\over 4 \sinh \xi_{-} \sinh \xi_{+} \sinh^{2N-1} \eta 
\cosh \eta} - {\sinh^{2}\eta  + N \cosh^{2}\eta\over 2 \cosh \eta} 
\id \,.
\ee
The corresponding energy eigenvalues $E$ are therefore given by
\be
E = {\Lambda'(0)\over 4 \sinh \xi_{-} \sinh \xi_{+} \sinh^{2N-1} \eta 
\cosh \eta} - {\sinh^{2}\eta  + N \cosh^{2}\eta\over 2 \cosh \eta}  \,,
\label{energy}
\ee
where $\Lambda(u)$ are eigenvalues of the transfer matrix.

The transfer matrix has the periodicity property
\be
t(u + i \pi)= t(u) \,,
\label{periodicity}
\ee
as follows from (\ref{periodicityR}), (\ref{periodicityK}). Moreover,
the transfer matrix has crossing symmetry
\be
t(-u - \eta)= t(u) \,,
\label{transfercrossing}
\ee
which can be proved using a generalization of the methods developed in
\cite{MN2} for the special case of quantum-group symmetry.  Finally,
we note that the transfer matrix has the asymptotic behavior (for
$\kk_{\pm} \ne 0$)
\be
t(u) \sim -\kk_{-}\kk_{+} {e^{u(2N+4)+\eta (N+2)}\over 2^{2N+1}} \id + 
\ldots \qquad \mbox{for} \qquad
u\rightarrow \infty \,.
\label{transfasympt}
\ee

\section{Fusion}\label{sec:fusion}

Our main tool is the so-called fusion technique, by which
higher-dimensional representations can be obtained from
lower-dimensional ones.  The fusion technique was first developed in
\cite{Ka, KRS, KS} for $R$ matrices, and then later generalized in
\cite{MNR} - \cite{IOZ} for $K$ matrices.  Following \cite{KRS}, we
introduce the (undeformed) projectors
\be
P^{\pm}_{1 \ldots m}={1\over m!}\sum_{\sigma}(\pm 1)^{\sigma} {\cal 
P}_{\sigma} \,,
\ee
where the sum is over all permutations $\sigma = (\sigma_{1} \,, 
\ldots \sigma_{m})$ of $(1\,, \ldots \,, m)$, and ${\cal P}_{\sigma}$ 
is the permutation operator in the space $\otimes_{k=1}^{m} {\cal 
C}^{2}$. For instance,
\be
P^{+}_{12} &=& {1\over 2}(\id + {\cal P}_{12}) \,, \non \\
P^{+}_{123} &=& {1\over 6}(\id + {\cal P}_{23}{\cal P}_{12}+{\cal 
P}_{12}{\cal P}_{23}+{\cal P}_{12}+{\cal P}_{23}+{\cal P}_{13}) \,.
\ee
The fused spin-$(j,{1\over 2})$ $R$ matrix ($j = {1\over 2} \,, 1 \,, 
{3\over 2}\,, \ldots $) is given by \cite{KRS, KS}
\be
R_{\langle 1 \ldots 2j \rangle 2j+1}(u) = P^{+}_{1 \ldots 2j} 
R_{1, 2j+1}(u) R_{2, 2j+1}(u+\eta) \ldots R_{2j, 2j+1}(u +(2j-1)\eta) 
P^{+}_{1 \ldots 2j} \,.
\label{fusedR}
\ee 
We also note
\be
R_{2j+1\langle 1 \ldots 2j \rangle}(u) &=& P^{+}_{1 \ldots 2j} 
R_{2j+1, 2j}(u -(2j-1)\eta) \ldots R_{2j+1, 2}(u-\eta) 
R_{2j+1, 1}(u) P^{+}_{1 \ldots 2j} \non  \\
&=& R_{\langle 1 \ldots 2j \rangle 2j+1}(u-(2j-1)\eta) \,.
\ee
The fused spin-$j$ $K^{-}$ matrix is given by \cite{MN1, Zh}
\be
K^{-}_{\langle 1 \ldots 2j \rangle}(u) &=& P^{+}_{1 \ldots 2j} 
K^{-}_{2j}(u) R_{2j, 2j-1}(2u+\eta) K^{-}_{2j-1}(u+\eta) \non \\
&\times & 
R_{2j,2j-2}(2u+2\eta) R_{2j-1,2j-2}(2u+3\eta) K^{-}_{2j-2}(u+2\eta) \non \\
&\times& \ldots
R_{2j,1}(2u+(2j-1)\eta) R_{2j-1,1}(2u+2j\eta) \ldots 
R_{2,1}(2u+(4j-3)\eta) \non \\
&\times& K^{-}_{1}(u+(2j-1)\eta) P^{+}_{1 \ldots 2j}  \,.
\label{fusedKm}
\ee
The fused spin-$j$ $K^{+}$ matrix is given by $K^{-}_{\langle 1 \ldots
2j \rangle}(-u- 2j\eta)$ with $(\xi_{-} \,, \kk_{-})$ replaced by
$(\xi_{+} \,, \kk_{+})$,
\be
K^{+}_{\langle 1 \ldots 2j \rangle}(u) = 
K^{-}_{\langle 1 \ldots 2j \rangle}(-u- 2j\eta) 
\Big\vert_{(\xi_{-} \,, \kk_{-}) \rightarrow (\xi_{+} \,, \kk_{+})}
\,. \label{fusedKp}
\ee
The fused (boundary) matrices satisfy 
generalized (boundary) Yang-Baxter equations.

The fused transfer matrix $t^{(j)}(u)$ constructed with a spin-$j$ 
auxiliary space is given by
\be
t^{(j)}(u)= \tr_{1 \ldots 2j} 
K^{+}_{\langle 1 \ldots 2j \rangle}(u) 
T_{\langle 1 \ldots 2j \rangle}(u) 
K^{-}_{\langle 1 \ldots 2j \rangle}(u) \hat T_{\langle 1 \ldots 2j 
\rangle}(u+(2j-1)\eta) \,,
\label{fusedtransfer}
\ee 
where 
\be
T_{\langle 1 \ldots 2j \rangle}(u) 
&=& R_{\langle 1 \ldots 2j \rangle N}(u) \ldots 
R_{\langle 1 \ldots 2j \rangle 1}(u) \,, \non \\
\hat T_{\langle 1 \ldots 2j \rangle}(u+(2j-1)\eta) 
&=& R_{\langle 1 \ldots 2j \rangle 1}(u) \ldots
R_{\langle 1 \ldots 2j \rangle N}(u) \,.
\ee 
The transfer matrix (\ref{transfer}) corresponds to the fundamental case 
$j={1\over 2}$; that is, $t^{({1\over 2})}(u) = t(u)$.
The fused transfer matrices constitute commutative families
\be
\left[ t^{(j)}(u)\,, t^{(k)}(v) \right] = 0  \,.
\ee 
These transfer matrices also satisfy a so-called fusion hierarchy 
\cite{MN1, Zh}
\be
t^{(j)}(u) &=& \tilde \zeta_{2j-1}(2u+ (2j-1) \eta) \Big[
t^{(j-{1\over 2})}(u)\ t^{({1\over 2})}(u+ (2j-1)\eta ) \non \\
&-& {\Delta(u+(2j-2)\eta)\ \tilde \zeta_{2j-2}(2u+ (2j-2) \eta)  \over 
\zeta(2u+ 2(2j-1)\eta)}\ t^{(j-1)}(u) \Big] \,, 
\label{hierarchy}
\ee
with $t^{(0)} = \id$, and $j= 1 \,, {3\over 2}\,, \ldots $. 
The quantity $\Delta (u)$, the so-called quantum determinant \cite{IK, KS},  
is given by
\be
\Delta (u) = \Delta \left\{ K^+(u) \right\} \Delta \left\{ K^-(u) \right\}
\delta \left\{ T(u) \right\} \delta \left\{ \hat T(u) \right\},
\label{qdeterminant}
\ee
where 
\be
\delta \left\{ T(u) \right \} &=& \tr_{12} \left\{ P_{12}^- \
T_1(u)\ T_2(u+\eta) \right\} = \zeta(u + \eta)^N \,, \non \\
\delta \left\{\hat T(u) \right\} &=&  \tr_{12} \left\{ P_{12}^- \
\hat T_2(u)\ \hat T_1(u+\eta) \right\} = \zeta(u + \eta)^N \,, \non \\
\Delta \left\{ K^-(u) \right\} &=& 
\tr_{12} \left\{  P_{12}^- \ K^-_1(u)\ 
R_{12}(2u + \eta)\ K^-_2(u+\eta)\ \right\}  \non \\
&=& -\sinh 2u \left[\sinh(u + \eta + \xi_{-}) 
\sinh(u + \eta - \xi_{-})
+ \kk_{-}^{2} \sinh^{2}(2u+2\eta) \right] \,, \non \\
\Delta \left\{ K^+(u) \right\} &=& 
\tr_{12} \left\{ P_{12}^- \ K^+_2 (u + \eta)\ 
R_{12}(-2u-3\eta)\  K^+_1(u) \right\}  \non \\
&=& \Delta \left\{ K^-(-u-2\eta) \right\} 
\Big\vert_{(\xi_{-} \,, \kk_{-}) \rightarrow (\xi_{+} \,, \kk_{+})} \,. 
\label{qdeterminants}
\ee
Moreover,
\be
\tilde \zeta_{j}(u) = \prod_{k=1}^{j} \zeta(u + k \eta) \,, 
\qquad \tilde \zeta_{0}(u) = 1 \,.
\ee

\section{Truncation identity}\label{sec:truncation}

We now proceed to formulate the important identity (\ref{truncation}),
which serves to truncate the fusion hierarchy (\ref{hierarchy}).  To
this end, we first derive separate ``truncation'' identities for the
$R$ and $K$ matrices.
 
\subsection{$R$ matrix truncation}\label{sec:Rtruncation}

We recall that, in addition to the fusion approach described
above, there is an alternative construction \cite{KR} of higher-spin
$R$ matrices based on quantum groups.  Following the notation of 
\cite{BGF}, the spin-$({1\over 2}\,, j)$ $R$ matrix is given by
\be
R_{({1\over 2}\,, j)}^{qg}(u) =
\left( \begin{array}{cc}
     \sinh \left(u+ ({1\over 2} + \hat H)\eta \right)  
     &\sinh \eta\ \hat F  \\
     \sinh \eta\ \hat E  
     & \sinh \left(u+ ({1\over 2} - \hat H)\eta \right)           
\end{array} \right) \,, 
\label{Rqg}
\ee
where the matrices $\hat H$, $\hat E$ and $\hat F$ have matrix elements
\be
(\hat H)_{mn} &=& (j+1-n) \delta_{m,n} \,, \qquad m \,, n = 1 \,, 2 \,, 
\ldots \,, 2j+1 \,, \non \\
(\hat E)_{mn} &=&  \omega_{m}\delta_{m,n-1} \,, \qquad
(\hat F)_{mn} = \omega_{n}\delta_{m-1,n} \,, \qquad 
\omega_{n}=\sqrt{[n]_{q}\ [2j+1-n]_{q}} \,,
\ee
and 
\be
[x]_{q} = {q^{x} - q^{-x}\over q - q^{-1}} \,, \qquad q=e^{\eta} \,.
\ee
These matrices form a $(2j+1)$-dimensional representation of the
$U_{q}(su(2))$ algebra
\be
[ \hat H \,, \hat E ] = \hat E \,, \qquad 
[ \hat H \,, \hat F ] = -\hat F \,, \qquad 
[ \hat E \,, \hat F ] = [2 \hat H]_{q} \,.
\ee
The corresponding spin-$(j\,, {1\over 2})$ $R$ matrix 
$R_{(j\,, {1\over 2})}^{qg}$ is then given by
\be
{}_{\alpha \beta} \left( R_{(j\,, {1\over 2})}^{qg}(u) 
\right){}_{\alpha' \beta'}
= {}_{\beta \alpha}\left( R_{({1\over 2}\,, j)}^{qg}(u) 
\right){}_{\beta' \alpha'}
\,.
\ee 
We refer to these $R$ matrices as ``quantum group'' ($qg$) $R$ matrices
in order to distinguish them from the fused $R$ matrices constructed
previously (\ref{fusedR}).  The two sets of $R$ matrices are related as 
follows \footnote{\label{prune}It is understood that the ``null'' rows
and columns (i.e., those with only zero matrix elements) are to be
pruned from the LHS. We have verified this relation explicitly for 
$j=1 \,, {3\over 2} \,, 2$, and we conjecture that it is true for 
all $j$.}
\be
B_{1 \ldots 2j} A_{1 \ldots 2j}\ R_{\langle 1 \ldots 2j \rangle 2j+1}(u)\ 
A_{1 \ldots 2j}^{-1} B_{1 \ldots 2j}^{-1}= 
\left[\prod_{k=1}^{2j-1} \sinh(u + k \eta)\right]  
R_{(j\,, {1\over 2})}^{qg}(u + (2j-1){\eta\over 2}) \,,
\label{Rrelation}
\ee
where $A_{1 \ldots 2j}$ is the matrix of (unnormalized) Clebsch-Gordon
coefficients in the decomposition of the tensor product of $2j$
spin-${1\over 2}$ representations into a direct sum of $su(2)$
irreducible representations.  Moreover, $B_{1 \ldots 2j}$ is a
$u$-independent diagonal matrix which renders symmetric the matrix on
the LHS of (\ref{Rrelation}). The $A$ and $B$ matrices for the cases 
$j=1 \,, {3\over 2} \,, 2$ are given in Appendix \ref{sec:similarity}.

The quantum-group $R$ matrices have a particularly simple 
``truncation'' property. Indeed, we find that for 
$\eta = {i \pi\over p+1}$, the spin-$({p+1\over 2}\,, {1\over 2})$
$R$ matrix takes the block-diagonal form
\be
R_{({p+1\over 2}\,, {1\over 2})}^{qg}(u) =
\left( \begin{array}{ccc}
  i\cosh(u + {\eta\over 2})\sigma^{z} &0           &0            \\
  0                         &R_{({p-1\over 2}\,, {1\over 2})}^{qg}(u)  &0 \\
  0                         &0           &-i\cosh(u + {\eta\over 2})\sigma^{z}
\end{array} \right) \,.
\ee
In view of the relation (\ref{Rrelation}), we see that 
the corresponding fused spin-$({p+1\over 2}\,, {1\over 2})$ $R$ matrix satisfies
\be
\lefteqn{B_{1 \ldots p+1} A_{1 \ldots p+1}\ 
R_{\langle 1 \ldots p+1 \rangle p+2}(u)\ 
A_{1 \ldots p+1}^{-1} B_{1 \ldots p+1}^{-1}} \non \\
&=& \mu(u) \left( \begin{array}{ccc}
  \nu(u) \sigma^{z} &0            &0    \\
  0                 
  &B_{1 \ldots p-1} A_{1 \ldots p-1}\ R_{\langle 1 \ldots p-1 \rangle 
    p}(u+\eta)\ A_{1 \ldots p-1}^{-1} B_{1 \ldots p-1}^{-1}    &0          \\
  0                 &0            &-\nu(u) \sigma^{z}           
\end{array} \right) \,, 
\ee
where 
\be
\mu(u) = \zeta(u) \,,
\label{mu}
\ee
and 
\be
\nu(u) = - {1\over \mu(u)}\prod_{k=0}^{p} \sinh(u + k \eta) 
= - {1\over \mu(u)} \left({i\over 2}\right)^{p} \sinh((p+1)u) \,.
\label{nu}
\ee
In obtaining this result, we have used the identity (see
1.392 in \cite{GR})
\be
\prod_{k=0}^{p} \sinh(u + k \eta) \Big\vert_{\eta={i \pi\over p+1}}
= \left({i\over 2}\right)^{p} \sinh((p+1)u) \,.
\label{identity}
\ee

As will be explained in Section \ref{sec:Ktruncation} below, it is
actually more useful to consider the similarity transformation with a
matrix $C$ (instead of $B$), which results in a triangular (instead of
a symmetric, block-diagonal) matrix.  For later convenience, we
present this result here:
\be
\lefteqn{C_{1 \ldots p+1} A_{1 \ldots p+1}\ 
R_{\langle 1 \ldots p+1 \rangle p+2}(u)\ 
A_{1 \ldots p+1}^{-1} C_{1 \ldots p+1}^{-1}} \non \\
&=& \mu(u) \left( \begin{array}{ccc}
  \nu(u) \sigma^{z} &0            &0    \\
  0                 
  &B_{1 \ldots p-1} A_{1 \ldots p-1}\ R_{\langle 1 \ldots p-1 \rangle 
    p}(u+\eta)\ A_{1 \ldots p-1}^{-1} B_{1 \ldots p-1}^{-1}    &*       \\
  0                 &0            &-\nu(u) \sigma^{z}           
\end{array} \right) \,. 
\label{Rtruncation}
\ee
The monodromy matrices therefore obey analogous relations
\be
\lefteqn{C_{1 \ldots p+1} A_{1 \ldots p+1}\ 
T_{\langle 1 \ldots p+1 \rangle}(u)\ 
A_{1 \ldots p+1}^{-1} C_{1 \ldots p+1}^{-1}} \non \\
&=& \mu(u)^{N} \left( \begin{array}{ccc}
  \nu(u)^{N} F &0            &0    \\
  0                 
  &B_{1 \ldots p-1} A_{1 \ldots p-1}\ 
  T_{\langle 1 \ldots p-1 \rangle}(u+\eta)\ 
  A_{1 \ldots p-1}^{-1} B_{1 \ldots p-1}^{-1}     &*        \\
  0                 &0            &(-\nu(u))^{N} F           
\end{array} \right) \,, \non \\
\lefteqn{C_{1 \ldots p+1} A_{1 \ldots p+1}\ 
\hat T_{\langle 1 \ldots p+1 \rangle}(u + p \eta )\ 
A_{1 \ldots p+1}^{-1} C_{1 \ldots p+1}^{-1}}\non \\
&=& \mu(u)^{N} \left( \begin{array}{ccc}
  \nu(u)^{N} F &0            &0    \\
  0                 
  &B_{1 \ldots p-1} A_{1 \ldots p-1}\ 
  \hat T_{\langle 1 \ldots p-1 \rangle}(u+(p-1)\eta)\ 
  A_{1 \ldots p-1}^{-1} B_{1 \ldots p-1}^{-1}      &*         \\
  0                 &0            &(-\nu(u))^{N} F           
\end{array} \right) \,,   \non \\
\label{monodromytruncation}
\ee
where $F=\prod_{k=1}^{N} \sigma^{z}_{k}$. The $C$ matrices for the cases 
$j=1 \,, {3\over 2} \,, 2$ are also given in Appendix \ref{sec:similarity}.

\subsection{$K$ matrix truncation}\label{sec:Ktruncation}

An explicit construction of nondiagonal higher-spin $K$ matrices
analogous to (\ref{Rqg}) is unfortunately not yet known. 
Nevertheless, considerable insight can be gained by first considering
the diagonal case (i.e., $\kk_{\pm}=0$).  The spin-$j$ diagonal $K^{-}$
matrix is given explicitly by (see \cite{MNR} for the case $j=1$)
\be
K^{- qg}_{(j)}(u)= diag \left( k_{(j)}^{(1)}(u) \,, k_{(j)}^{(2)}(u) \,, \ldots 
\,, k_{(j)}^{(2j+1)}(u) \right) \,,
\label{Kqg}
\ee
where
\be
k_{(j)}^{(m)}(u) = \prod_{l=0}^{2j-m} \sinh(\xi_{-} + u + l \eta)
 \prod_{l=0}^{m-2} \sinh(\xi_{-} - u - l \eta) \,.
\ee
This $K$ matrix is related to the fused $K$ matrix (\ref{fusedKm}) 
by \footnote{Footnote \ref{prune} applies here too.}
\be
A_{1 \ldots 2j}\ K^{-}_{\langle 1 \ldots 2j \rangle}(u)\ 
A_{1 \ldots 2j}^{-1} = \left[ \prod_{l=1}^{2j-1} \prod_{k=1}^{l}
\sinh(2u + (l+k)\eta)\right]  K_{(j)}^{-qg}(u) \,.
\label{Krelation}
\ee

We find that the diagonal $K^{- qg}_{(j)}(u)$ with $j={p+1\over 2}$
has the following ``truncation'' property for 
$\eta = {i \pi\over p+1}$,
\be
\lefteqn{K^{- qg}_{({p+1\over 2})}(u) = 
\sinh(\xi_{-} + u) \sinh(\xi_{-} - u)} \non \\
&\times & \left( \begin{array}{ccc}
\left({i\over 2}\right)^{p} 
{\sinh((p+1)(\xi_{-}+u))\over  \sinh(\xi_{-} + u) \sinh(\xi_{-} - u)} &0          
&0            \\
0           &K^{- qg}_{({p-1\over 2})}(u+\eta)   &0 \\
0           &0           &\left(-{i\over 2}\right)^{p} 
{\sinh((p+1)(\xi_{-}-u))\over  \sinh(\xi_{-} + u) \sinh(\xi_{-} - u)}
\end{array} \right) \,.
\ee
It follows that for $\eta = {i \pi\over p+1}$, the diagonal fused
spin-${p+1\over 2}$ $K^{-}$ matrix satisfies the truncation identity
\be
\lefteqn{A_{1 \ldots p+1}\ K_{\langle 1 \ldots p+1 \rangle}^{-}(u)\ 
A_{1 \ldots p+1}^{-1}} \non \\
&=& \mu_{-}(u) \left( \begin{array}{ccc}
	\nu_{-}(u)  &0            &0   \\
0             
&A_{1 \ldots p-1}\ K_{\langle 1 \ldots p-1 \rangle}^{-}(u+\eta)\ 
	A_{1 \ldots p-1}^{-1}             &0          \\
	0   &0           &\nu'_{-}(u)           
\end{array} \right) \,, 
\label{diagKtruncation}
\ee
where
\be
\mu_{-}(u) &=& {\Delta\{ K^{-}(u-\eta)\}\over \sinh(2u-2\eta)}
\prod_{k=2}^{2p}\sinh(2u+ k \eta) \non  \\
 &=&  -{\Delta\{ K^{-}(u-\eta)\} \sinh^{2}\left(2(p+1)u\right)
\over 2^{2p}\sinh 2u \sinh(2u+\eta)\sinh(2u-\eta)\sinh(2u-2\eta)}
\,,
\label{muM}
\ee
the quantum determinant $\Delta\{ K^{-}(u)\}$ is given in Eq. 
(\ref{qdeterminants}),
\be 
\nu_{-}(u) &=& {1\over \mu_{-}(u)} \left({i\over 2}\right)^{p} 
\left[  \prod_{l=1}^{p} \prod_{k=1}^{l} \sinh(2u + (l+k)\eta) \right]
\sinh \left( (p+1)(\xi_{-}+u) \right) \non  \\
&=& {1\over \mu_{-}(u)} 
{e^{{1\over 2}i \pi p(p+2)}\over 2^{{1\over 2}p(p+1)}}
\cosh^{[{p\over 2}]}((p+1)u) \sinh^{[{p+1\over 2}]}((p+1)u) \non  \\
&\times& \sinh \left( (p+1)(\xi_{-}+u) \right) \,,
\label{nuMdiag}  
\ee 
and
\be 
\nu'_{-}(u) = \pm \nu_{-}(u) \qquad  \mbox{with} \qquad 
\xi_{-} \rightarrow -\xi_{-} \,. \label{nuMprime}
\ee
The standard notation $[x]$ denotes integer part of $x$.  

We turn now to the more general nondiagonal case ($\kk_{\pm} \ne 0$). 
The boundary matrices
\be
B_{1 \ldots 2j}A_{1 \ldots 2j}\ K_{\langle 1 \ldots 2j \rangle}^{\mp}(u)\ 
A_{1 \ldots 2j}^{-1} B_{1 \ldots 2j}^{-1} \,, \non 
\ee
like their bulk counterpart (\ref{Rrelation}), are symmetric.  However,
for $\eta = {i \pi\over p+1}$ and $j={p+1\over 2}$, these matrices are not
block-diagonal.  In order to obtain a truncation identity for the full
transfer matrix, it is desirable to have at least block-triangular
matrices.  

To this end, we consider a new similarity transformation, replacing
the matrix $B$ by a new matrix $C$ which is also diagonal and
$u$-independent.  (See Appendix \ref{sec:similarity} for explicit
expressions of the $C$ matrices for $j=1 \,, {3\over 2} \,, 2$.) Indeed, 
we propose the following generalization 
of (\ref{diagKtruncation}) for the nondiagonal case: \footnote{This equation 
is formal, since $\sigma_{\mp} \rightarrow \infty$ 
(and $\rho_{\mp} \rightarrow 0$) 
in the limit  $\eta \rightarrow {i \pi\over p+1}$, as can be seen
from Eqs. (\ref{sigmaMrhoM}) and (\ref{afactor}). However, substituting 
this result into the expression for the fused transfer matrix gives a 
final result (\ref{truncationagain}) which is finite
in the $\eta \rightarrow {i \pi\over p+1}$ limit.}
\be
\lefteqn{C_{1 \ldots p+1} A_{1 \ldots p+1}\ 
K_{\langle 1 \ldots p+1 \rangle}^{\mp}(u)\ 
A_{1 \ldots p+1}^{-1} C_{1 \ldots p+1}^{-1}} \non \\
&=& \mu_{\mp}(u) \left( \begin{array}{ccc}
	\nu_{\mp}(u)  &0            &\sigma_{\mp}(u)   \\
0             
&B_{1 \ldots p-1} A_{1 \ldots p-1}\ 
K_{\langle 1 \ldots p-1 \rangle}^{\mp}(u+\eta)\ 
A_{1 \ldots p-1}^{-1} B_{1 \ldots p-1}^{-1}    &*          \\
	\rho_{\mp}(u)    &0           &\nu'_{\mp}(u)           
\end{array} \right) \,, 
\label{Ktruncation}
\ee
where $\mu_{-}(u)$ and $\nu'_{-}(u)$ are given by the same expressions 
(\ref{muM}), (\ref{nuMprime}); but the expression (\ref{nuMdiag}) 
for $\nu_{-}(u)$ is now replaced by
\be 
\nu_{-}(u) &=& {1\over \mu_{-}(u)} 
{e^{{1\over 2}i \pi p(p+2)}\over 2^{{1\over 2}p(p+1)}}
\cosh^{[{p\over 2}]}((p+1)u) \sinh^{[{p+1\over 2}]}((p+1)u)\ 
n(u \,; \xi_{-} \,, \kk_{-})
\,,
\label{nuM}
\ee
where the function $n(u \,; \xi \,, \kk)$ is defined by
\be
n(u \,; \xi \,, \kk) = \sinh \left( (p+1)(\xi +u) \right)  
+ \sum_{l=1}^{\left[{p+1\over 2}\right]}c_{p\,, l}\ 
\kk^{2l} \sinh \left( (p+1)u + (p+1 - 2l) \xi \right) \,.
\label{nfunction}
\ee 
We have explicitly computed
the coefficients $c_{p\,,l}$ which appear in this function for values
of $p$ up to $p=5$, and we find that they are consistent with the
following formulas:
\be
c_{p \,, 1} &=& p + 1 \,, \non \\
c_{p \,, 2} &=& {1\over 2}p(p-1) -1 \,,
\label{ccoefficients}
\ee
and also $c_{5 \,, 3}= 2$. It remains a challenge to determine
the coefficients $c_{p\,,l}$ for all values of $p$ and $l$. Moreover, 
\be
\sigma_{-}(u) =  {a\ \omega_{-}(u)\over \mu_{-}(u)} \,, \qquad
\rho_{-}(u) = {\omega_{-}(u)\over  a\ \mu_{-}(u)} \,,
\label{sigmaMrhoM}
\ee
where
\be
\omega_{-}(u) &=& \kk_{-}^{p+1} \left[ \prod_{l=1}^{p} 
\prod_{k=1}^{l} \sinh(2u + (l+k-1)\eta) \right] \prod_{k=0}^{p}
\sinh(2u+ 2k \eta) \non  \\
 &=& {e^{{1\over 2}i \pi p(p+2)}\over 2^{{1\over 2}p(p+1)-1}} 
\kk_{-}^{p+1} \cosh^{[{p+2\over 2}]}((p+1)u) 
\sinh^{[{p+3\over 2}]}((p+1)u) \,, 
\label{omegaM}
\ee
and
\be 
a = \left( [p+1]_{q} \right)^{-{1\over 2}} \,.
\label{afactor}
\ee
Note that $a \rightarrow \infty$ for $\eta \rightarrow {i \pi\over p+1}$.

The $K^{+}$ matrices are given, in view of Eq. (\ref{fusedKp}) 
with $\eta= {i \pi\over p+1}$ and $j={p+1\over 2}$, by
\be
K^{+}_{\langle 1 \ldots p+1 \rangle}(u) = 
K^{-}_{\langle 1 \ldots p+1 \rangle}(-u- i \pi) 
\Big\vert_{(\xi_{-} \,, \kk_{-}) \rightarrow (\xi_{+} \,, \kk_{+})}
\,.
\ee
Hence, the ``plus'' quantities $\mu_{+}$, $\nu_{+}$, etc. can be 
readily obtained from the corresponding ``minus'' quantities 
$\mu_{-}$, $\nu_{-}$, etc. by making the replacements 
$u \rightarrow -u- i \pi$ and 
$(\xi_{-} \,, \kk_{-}) \rightarrow (\xi_{+} \,, \kk_{+})$.

\subsection{Transfer matrix truncation}\label{sec:transftruncation}

We are finally in position to formulate the truncation identity for
the fused transfer matrix $t^{(j)}(u)$ defined in
(\ref{fusedtransfer}).  Recalling the results
(\ref{monodromytruncation}) for the monodromy matrices
and (\ref{Ktruncation}) for the $K$ matrices, we obtain (for 
$\eta = {i \pi\over p+1}$ and $j={p+1\over 2}$)
\be
t^{({p+1\over 2})}(u) = \alpha(u) \left[ t^{({p-1\over 2})}(u+ \eta) + 
\beta(u) \id \right] \,,
\label{truncationagain}
\ee
where
\be
\alpha(u) &=& \mu(u)^{2N} \mu_{-}(u) \mu_{+}(u) \,, \non \\
\beta(u) &=& \nu(u)^{2N} \Big[\nu_{-}(u) \nu_{+}(u) +
\nu'_{-}(u) \nu'_{+}(u) \non \\
&+& (-1)^{N} \left( \sigma_{+}(u) \rho_{-}(u) + 
\sigma_{-}(u) \rho_{+}(u) \right) \Big] \,.
\label{alphabeta}
\ee
Note that the factors of $a$ from $\sigma_{\mp}(u)$ and
$\rho_{\mp}(u)$ cancel in the expression for $\beta(u)$; and hence,
the result is finite for $\eta \rightarrow {i \pi\over p+1}$.

\section{Functional relations}\label{sec:functional}

Combining the fusion hierarchy (\ref{hierarchy}) and the truncation
identity (\ref{truncationagain}), it is straightforward to obtain -- for
any positive integer value of $p$ -- a $(p+1)$-order functional relation
for the fundamental transfer matrix. We propose the following general 
form of the functional relations:
\be
\lefteqn{f_{0}(u) t(u) t(u +\eta) \ldots t(u + p \eta)} \non \\
&-& f_{1}(u) t(u +\eta) t(u +2\eta) \ldots t(u + (p-1)\eta) \non \\
&-& f_{1}(u+\eta) t(u +2\eta) t(u +3\eta)\ldots t(u + p \eta) \non \\
&-& f_{1}(u+2\eta) t(u) t(u +3\eta) t(u +4\eta) \ldots t(u + p \eta) \non \\
&-& f_{1}(u+3\eta) t(u) t(u +\eta) t(u +4\eta) 
\ldots t(u + p \eta) - \ldots \non \\
&-& f_{1}(u+p\eta) t(u) t(u +\eta) \ldots t(u +  (p-2)\eta) \non \\
&+& {f_{1}(u) f_{1}(u +2\eta)\over f_{0}(u)}  
t(u +3\eta) t(u +4\eta) \ldots t(u + (p-1)\eta) \non \\
&+& {f_{1}(u) f_{1}(u +3\eta)\over f_{0}(u)} 
t(u +\eta) t(u +4\eta) t(u +5\eta)\ldots t(u + (p-1)\eta) + \ldots \non \\
&+& {f_{1}(u) f_{1}(u +(p-1)\eta)\over f_{0}(u)}  
t(u +\eta) t(u +2\eta) \ldots t(u + (p-3)\eta) \non\\
&+& {f_{1}(u+\eta) f_{1}(u +3\eta)\over f_{0}(u)}  
t(u +4\eta) t(u +5\eta) \ldots t(u + p\eta) \non \\
&+& {f_{1}(u+\eta) f_{1}(u +4\eta)\over f_{0}(u)} t(u +2\eta) t(u +5\eta) 
t(u +6\eta)\ldots t(u + p\eta) + \ldots \non \\
&+& {f_{1}(u+\eta) f_{1}(u +p\eta)\over f_{0}(u)}  
t(u +2\eta) t(u +3\eta) \ldots t(u + (p-2)\eta) \non \\
&+& {f_{1}(u+2\eta) f_{1}(u +4\eta)\over f_{0}(u)}  
t(u) t(u +5\eta) t(u +6\eta) \ldots t(u + p\eta) \non \\
&+& {f_{1}(u+2\eta) f_{1}(u +5\eta)\over f_{0}(u)} t(u) t(u +3\eta) 
t(u +6\eta) t(u +7\eta)\ldots t(u + p\eta) + \ldots \non \\
&+& {f_{1}(u+2\eta) f_{1}(u +p\eta)\over f_{0}(u)}  
t(u) t(u +3\eta) t(u +4\eta)\ldots t(u + (p-2)\eta)  \non \\
&+& \ldots  = f_{3}(u) \,.
\label{funcrltn1}
\ee 
In particular, the first three functional relations are given by
\be
&p=1:& \qquad f_{0}(u) t(u) t(u +\eta) - f_{1}(u) - f_{1}(u+\eta)  = 
f_{3}(u) \,, \non  \\
&p=2:& \qquad f_{0}(u) t(u) t(u +\eta) t(u +2\eta) - f_{1}(u) t(u +\eta) 
- f_{1}(u+\eta) t(u +2\eta) \non  \\
& & \qquad \qquad - f_{1}(u +2\eta) t(u)   = f_{3}(u) \,, \non  \\
&p=3:& \qquad f_{0}(u) t(u) t(u +\eta) t(u +2\eta) t(u +3\eta) 
- f_{1}(u) t(u +\eta) t(u +2\eta) \non  \\
& & \qquad \qquad - f_{1}(u+\eta) t(u +2\eta) t(u +3\eta) 
- f_{1}(u +2\eta) t(u) t(u +3\eta)   \non  \\
& & \qquad \qquad 
- f_{1}(u +3\eta) t(u) t(u +\eta)  = f_{3}(u) \,.
\ee 

Remarkably, only three distinct functions $f_{0}$, $f_{1}$, $f_{3}$
appear in the functional relations.  The function $f_{0}(u)$ is given
by
\be
f_{0}(u) &=&  \prod_{l=1}^{p} \prod_{k=1}^{l} 
\zeta \left( 2u + (l+k) \eta \right) \non \\
&=& {e^{i {\pi\over 2}(p+1)^{2}}\over 2^{p(p-1)}} \cosh^{p}((p+1)u) 
\sinh^{p-1}((p+1)u) \cosh ((p+1)(u - {i\pi\over 2})) \,.
\label{f0}
\ee 
This function has the periodicity property
\be
f_{0}(u) = f_{0}(u + \eta) 
\ee
for $\eta = {i \pi\over p+1}$. The function $f_{1}(u)$ is given by
\be
f_{1}(u) = {f_{0}(u) \Delta(u - \eta)\over \zeta(2u)} \,,
\label{f1}
\ee 
where $\zeta(u)$ is given by (\ref{unitarity}), and the quantum
determinant $\Delta(u)$ is given by (\ref{qdeterminant}).
Finally, the function $f_{3}(u)$ is given by
\be
f_{3}(u) &=& \alpha(u) \beta(u) \non \\
&=& {(-1)^{N p + [{p+1\over 2}]}\over 2^{p(p+1+2N)}} 
\cosh^{2[{p\over 2}]}((p+1)u) \sinh^{2N+ 2[{p+1\over 2}]}((p+1)u)  \non \\
&\times& \Big\{ 
n(u \,; \xi_{-} \,, \kk_{-})\ n(u \,; -\xi_{+} \,, \kk_{+}) +
n(u \,; -\xi_{-} \,, \kk_{-})\ n(u \,; \xi_{+} \,, \kk_{+}) \non \\
&\quad&+ 2 (-1)^{N} (-\kk_{-} \kk_{+})^{p+1} \sinh^{2}(2(p+1)u) \Big\} \,,
\label{f3}
\ee 
where the function $n(u \,; \xi \,, \kk)$ is defined in Eq. 
(\ref{nfunction}).  As already mentioned, we have computed the
coefficients $c_{p\,, l}$ which appear in this function only up to
$p=5$. 

We remark that, in obtaining the result (\ref{funcrltn1}), we have 
made use of the relation
\be
f_{1}(u) = \alpha(u) \prod_{l=1}^{p-2} \prod_{k=1}^{l} 
\zeta \left( 2u + (l+k+2) \eta \right) 
\ee 
which is satisfied by the function $f_{1}(u)$ defined in (\ref{f1}).

\section{Eigenvalues and Bethe Ansatz equations}\label{sec:eigenvalues}

The functional relations which we have obtained can be used to determine the 
eigenvalues of the transfer matrix.  The commutativity relation 
(\ref{commutativity}) implies that the transfer matrix has eigenstates 
$| \Lambda \rangle$ which are independent of $u$,
\be
t(u) |\Lambda \rangle = \Lambda(u) | \Lambda \rangle \,,
\label{eigenvalueproblem}
\ee
where $\Lambda(u)$ are 
the corresponding eigenvalues.  Acting on $|\Lambda \rangle$ 
with the functional relation (\ref{funcrltn1}), we obtain the 
corresponding relation for the eigenvalues
\be
&& f_{0}(u) \Lambda(u) \Lambda(u +\eta) 
\ldots \Lambda(u + p \eta)
- f_{1}(u) \Lambda(u +\eta) \Lambda(u +2\eta) 
\ldots \Lambda(u + (p-1)\eta) \non  \\
&& \qquad + \ldots  = f_{3}(u) \,.
\label{funcrltn2}
\ee
Similarly, it follows from (\ref{periodicity}) and
(\ref{transfercrossing}) that the eigenvalues have the periodicity and
crossing properties
\be
\Lambda(u + i \pi) = \Lambda(u) \,, \qquad 
\Lambda(-u - \eta) = \Lambda(u) \,.
\label{eigenvalueprops}
\ee
Finally, (\ref{transfasympt}) implies the asymptotic behavior (for
$\kk_{\pm} \ne 0$)
\be
\Lambda(u) \sim -\kk_{-}\kk_{+} {e^{u(2N+4)+\eta (N+2)}\over 2^{2N+1}} + 
\ldots \qquad \mbox{for} \qquad
u\rightarrow \infty \,.
\label{asympt}
\ee

We shall assume that the eigenvalues have the form
\be
\Lambda(u) = \rho \prod_{j=-1}^{N}\sinh(u - u_{j}) \sinh(u + \eta + u_{j}) 
\label{Ansatz} \,,
\ee
where $u_{j}$ and $\rho$ are ($u$-independent) parameters which are to 
be determined.  Indeed, this expression satisfies the periodicity and 
crossing properties (\ref{eigenvalueprops}), and it has the correct 
asymptotic behavior (\ref{asympt}) provided that we set
\be
\rho = -8 \kk_{-}\kk_{+} \,.
\label{rho}
\ee 

Evaluating the functional relation (\ref{funcrltn2}) at the root 
$u=u_{j}$, we obtain a set of Bethe-Ansatz-like equations
\be
\lefteqn{-f_{1}(u_{j}) \Lambda(u_{j}+\eta) \Lambda(u_{j}+2\eta) \ldots 
\Lambda(u_{j}+ (p-1)\eta)} \non \\
&+& {f_{1}(u_{j}) f_{1}(u_{j}+2\eta)\over f_{0}(u_{j})}  
\Lambda(u_{j}+3\eta) \Lambda(u_{j}+4\eta) \ldots 
\Lambda(u_{j}+ (p-1)\eta) \non \\
&+& {f_{1}(u_{j}) f_{1}(u_{j}+3\eta)\over f_{0}(u_{j})}  
\Lambda(u_{j}+\eta) \Lambda(u_{j}+4\eta) \Lambda(u_{j}+5\eta)\ldots 
\Lambda(u_{j}+ (p-1)\eta) + \ldots \non \\
&+& {f_{1}(u_{j}) f_{1}(u_{j}+(p-1)\eta)\over f_{0}(u_{j})}  
\Lambda(u_{j}+\eta) \Lambda(u_{j}+2\eta) \ldots 
\Lambda(u_{j}+ (p-3)\eta) + \ldots \non \\
&+& \left[u_{j} \rightarrow u_{j} + \eta \right] = f_{3}(u_{j}) \,,
\qquad j = -1 \,, 0 \,, \ldots \,, N \,,
\label{BAE1}
\ee 
where $\Lambda(u)$ is given by (\ref{Ansatz}).
In particular, the first three cases are given by
\be
&p=1:& \qquad f_{1}(u_{j}) + f_{1}(u_{j}+\eta) + f_{3}(u_{j}) = 0\,, \non  \\
&p=2:& \qquad f_{1}(u_{j}) \Lambda(u_{j} +\eta) 
+ f_{1}(u_{j}+\eta) \Lambda(u_{j} +2\eta) + f_{3}(u_{j}) = 0\,, \non  \\
&p=3:& \qquad f_{1}(u_{j}) \Lambda(u_{j} +\eta) \Lambda(u_{j} +2\eta) 
+ f_{1}(u_{j}+\eta) \Lambda(u_{j} +2\eta) \Lambda(u_{j} +3\eta) \non  \\
& & \qquad \qquad + f_{3}(u_{j}) = 0\,.
\ee 
The simplest case $p=1$ has recently been analyzed in 
\cite{NeXX}.\footnote{The functions $g_{1}(u)$ and $g_{3}(u)$ in 
\cite{NeXX} are related to $f_{1}(u)$ and $f_{3}(u)$ as follows:
\be
f_{1}(u) = -\sinh^{2} 2u\  \cosh^{4N}u\ g_{1}(u) \,, \qquad
f_{3}(u) = -\sinh^{2} 2u\  \sinh^{2N}u\ \cosh^{2N}u\ g_{3}(u) \,.
\non 
\ee}

The Bethe Ansatz equations may be written in a more explicit form by
substituting the Ansatz (\ref{Ansatz}) into (\ref{BAE1}), and then
using the identity (\ref{identity}) to simplify the products.  In this
way, we obtain
\be
\lefteqn{(-1)^{N+3}\left({i\over 2}\right)^{2p(N+2)} \rho^{p-1}}
\non  \\
&\times&\Big\{ f_{1}(u_{j}) \prod_{k=-1}^{N} 
{\sinh((p+1)(u_{j}-u_{k}))\ \sinh((p+1)(u_{j}+u_{k}))\over
\sinh(u_{j}-u_{k}) \sinh(u_{j}-u_{k}-\eta) \sinh(u_{j}+u_{k}+\eta)
\sinh(u_{j}+u_{k})} \non  \\
&+&f_{1}(u_{j}+\eta) \prod_{k=-1}^{N} 
{\sinh((p+1)(u_{j}-u_{k}+\eta))\ \sinh((p+1)(u_{j}+u_{k}+\eta))\over
\sinh(u_{j}-u_{k}+\eta) \sinh(u_{j}-u_{k}) \sinh(u_{j}+u_{k}+2\eta)
\sinh(u_{j}+u_{k}+\eta)} \Big\} \non \\
&+& \ldots = f_{3}(u_{j}) \,, \qquad j = -1 \,, 0 \,, \ldots \,, N \,.
\label{BAE2}
\ee
The poles which occur when $j=k$ are harmless, as they are canceled 
by corresponding zeros.

We conclude this section with a brief discussion of the diagonal case,
$\kk_{\pm} = 0$. In this case, the transfer matrix commutes 
with the matrix ${\cal M}$,
\be
\left[ t(u) \,, {\cal M} \right] = 0 \,,
\ee
where
\be
{\cal M} = {N\over 2} - S^{z} \,, \qquad S^{z} =
{1\over 2}\sum_{i=1}^{N} \sigma^{z}_{i} \,.
\ee
The two matrices can therefore be simultaneously diagonalized,
\be
t(u) |\Lambda^{(m)} \rangle &=& \Lambda^{(m)}(u) | \Lambda^{(m)} 
\rangle \,, \non \\
{\cal M} |\Lambda^{(m)} \rangle &=& m | \Lambda^{(m)} \rangle \,. 
\ee
The asymptotic behavior of the eigenvalues is now given by
\be
 \Lambda^{(m)}(u) \sim {i\over 2^{2N+2}} e^{u(2N+2)} \left(
e^{2(N-m)\eta + \xi_{-} - \xi_{+}} +
e^{2m\eta - \xi_{-} + \xi_{+}} \right) +
\ldots \qquad \mbox{for} \qquad
u\rightarrow \infty \,.
\ee
The eigenvalues are therefore given by
\be
\Lambda^{(m)}(u) = \rho^{(m)} \prod_{j=0}^{N}\sinh(u - u_{j}) 
\sinh(u + \eta + u_{j}) \,,
\label{diagAnsatz}
\ee
where
\be
\rho^{(m)} = 2i e^{-\eta} 
\cosh \left( \xi_{-}-\xi_{+} + (N - 2m)\eta \right) \,,
\ee 
and the roots $u_{j}$ satisfy essentially the same
Bethe Ansatz equations (\ref{BAE1}).

\section{Discussion}\label{sec:discuss}

We have demonstrated an approach for solving the open XXZ quantum spin
chain with nondiagonal boundary terms for $\eta = {i \pi\over p+1}$. 
In particular, we have proposed exact $(p+1)$-order functional
relations (\ref{funcrltn1}) for the transfer matrix, and Bethe Ansatz
equations (\ref{BAE1}), (\ref{BAE2}) for the corresponding
eigenvalues.  We emphasize that the function $f_{3}(u)$ appearing in
these equations involves the coefficients $c_{p \,, l}$ which we have
computed only up to $p=5$.  (See Eqs.  (\ref{f3}), (\ref{nfunction}),
(\ref{ccoefficients}).)  The complete determination of these
coefficients must presumably await a more explicit construction of
nondiagonal $K$ matrices for arbitrary spin.
 
A similar approach can certainly also be applied to the elliptic case
(i.e., the open XYZ chain with nondiagonal boundary terms), since both
a fusion hierarchy and a truncation identity can also be obtained for
this case.  Moreover, we expect that a similar approach can be applied
to spin chains constructed with $R$ and $K$ matrices associated with
any affine Lie algebra.

An important feature of our Bethe Ansatz equations is that, as in the
case of the closed (periodic boundary conditions) XYZ chain, there is a
fixed number of roots for all the eigenvalues.  Because of this fact,
it is not easy to compare our results in the diagonal limit
($\kk_{\pm} \rightarrow 0$) with those of \cite{ABBBQ, Sk}.  A similar
difficulty arises \cite{DDG} when comparing the Bethe Ansatz equations
of the closed XYZ chain in the trigonometric limit with the
``conventional'' Bethe Ansatz equations of the closed XXZ chain.

It should be interesting to use our results to determine, for low
values of $p$, physical properties (thermodynamics, etc.)  of the spin
chains and of the associated quantum field theories.  It would also be
interesting to try to further simplify our system of Bethe Ansatz
equations, and perhaps to generalize to the case of generic values of
the bulk anisotropy.

\section*{Acknowledgments}

I am grateful to M. Jimbo and Y. Yamada for helpful correspondence.
This work was supported in part by the National Science Foundation
under Grants PHY-9870101 and PHY-0098088.

\appendix

\section{Similarity transformations}\label{sec:similarity}

We discuss in Section \ref{sec:truncation} certain similarity
transformations on the fused $R$ and $K$ matrices.  Here we give
explicit expressions for the matrices $A \,, B \,, C$ corresponding to
these similarity transformations for the first three cases, namely $j=
1 \,, {3\over 2} \,, 2$.

For $j=1$,
\be
A =
\left( \begin{array}{cccc}
          1 &0          &0           &0 \\
	  0 &{1\over 2} &{1\over 2}  &0 \\
          0 &0          &0           &1 \\
	  0 &{1\over 2} &-{1\over 2} &0
\end{array} \right) \,, 
\ee
\be 
B= diag(a \,, 1 \,, a \,, 1) \,, \qquad 
C= diag(a \,, 1 \,, 1 \,, 1) \,, \qquad a = 
\left([2]_{q}\right)^{-{1\over 2}} \,.
\ee

For $j={3\over 2}$,
\be
A =
\left( \begin{array}{cccccccc}
 1 &0          &0           &0   &0          &0     &0   &0 \\
 0 &{1\over 3} &{1\over 3}  &0   &{1\over 3} &0     &0   &0 \\
 0 &0          &0   &{1\over 3} &0 &{1\over 3}   &{1\over 3} &0 \\
 0 &0          &0           &0   &0          &0     &0   &1\\
 0 &-{2\over 3} &{1\over 3} &0 &{1\over 3} &0     &0   &0\\
 0 &0          &0 &{1\over 3} &0 &{1\over 3} &-{2\over 3} &0\\
 0 &0 &{1\over 2} &0 &-{1\over 2} &0     &0   &0\\
 0 &0          &0 &{1\over 2} &0 &-{1\over 2} &0     &0 
\end{array} \right) \,, 
\ee
\be 
B= diag(a \,, 1 \,, 1 \,, a \,, 1\,, 1\,, 1\,, 1) \,, \qquad 
C= diag(a \,, 1 \,, \ldots \,, 1) \,, \quad a = 
\left([3]_{q}\right)^{-{1\over 2}} \,.
\ee

For $j=2$,
\be
A =
\left( \begin{array}{cccccccccccccccc}
    1 & 0 & 0 & 0 & 0 & 0 & 0 & 0 & 0 & 0 & 0 & 0 & 0 & 0 & 0 & 0  \\
    0 & {1\over 4} & {1\over 4} & 0 & {1\over 4} & 0 & 0 & 0 & {1\over 4}
     & 0 & 0 & 0 & 0 & 0 & 0 & 0  \\
    0 & 0 & 0 & {1\over 6} & 0 & {1\over 6} & {1\over 6} & 0 & 0 
    &{1\over 6} & {1\over 6} & 0 & {1\over 6} & 0 & 0 & 0  \\
    0 & 0 & 0 & 0 & 0 & 0 & 0 & {1\over 4} & 0 & 0 & 0 & {1\over 4} & 0 & 
    {1\over 4} & {1\over 4} & 0  \\
    0 & 0 & 0 & 0 & 0 & 0 & 0 & 0 & 0 & 0 & 0 & 0 & 0 & 0 & 0 & 1  \\
    0 & -{3\over 4} & {1\over 4} & 0 & {1\over 4} & 0 & 0 & 0 & {1\over 4} 
    & 0 & 0 & 0 & 0 & 0 & 0 & 0  \\
    0 & 0  & 0 & {1\over 6} & 0 & {1\over 6} & -{1\over 6} & 0 & 0 
    & {1\over 6}  
    & -{1\over 6} & 0 & -{1\over 6} & 0 & 0 & 0  \\
    0 & 0 & 0 & 0 & 0 & 0 & 0 & {1\over 4} & 0 & 0 & 0 & {1\over 4} & 
    0 & {1\over 4} & -{3\over 4} & 0  \\
    0 & 0 & -{2\over 3} & 0 & {1\over 3} & 0 & 0 & 0 & {1\over 3} & 0 & 
    0 & 0 & 0 & 0 & 0 & 0  \\
    0 & 0 & 0 & {1\over 3} & 0 & -{1\over 6} & {1\over 6} & 0 & 0 & 
    -{1\over 6} & {1\over 6} & 0 & -{1\over 3} & 0 & 0 & 0  \\
    0 & 0 & 0 & 0 & 0 & 0 & 0 & {1\over 3} & 0 & 0 & 0 & {1\over 3} & 0 & 
    -{2\over 3} & 0 & 0  \\
    0 & 0 & 0 & -{1\over 3} & 0 & {1\over 6} & {1\over 6} & 0 & 0 & {1\over 
    6} & {1\over 6} & 0 & -{1\over 3} & 0 & 0 & 0  \\
    0 & 0 & 0 & 0 & {1\over 2} & 0 & 0 & 0 & -{1\over 2} & 0 & 0 & 0 & 
    0 & 0 & 0 & 0  \\
    0 & 0 & 0 & 0 & 0 & {1\over 4} & {1\over 4} & 0 & 0 & -{1\over 4} 
    & -{1\over 4} & 0 & 0 & 0 & 0 & 0  \\
    0 & 0 & 0 & 0 & 0 & 0 & 0 & {1\over 2} & 0 & 0 & 0 & -{1\over 2} 
    & 0 & 0 & 0 & 0  \\
    0 & 0 & 0 & 0 & 0 & -{1\over 4} & {1\over 4} & 0 & 0 & {1\over 4} 
    & -{1\over 4} & 
    0 & 0 & 0 & 0 & 0
\end{array} \right) \,, 
\ee
\be 
B &=& diag(a \,, 1 \,, b \,, 1 \,, a \,, 1\,,  \ldots \,, 1) \,, \qquad 
C= diag(a \,, 1 \,, b \,, 1  \,,  \ldots \,, 1) \,, \non \\
a &=& \left([4]_{q}\right)^{-{1\over 2}} \,, \qquad  
b = \left({[2]_{q}\over [3]_{q}}\right)^{-{1\over 2}} \,.
\ee

\noindent
{\bf Note added:}

Boundary quantum group symmetry \cite{MN3, DM} can be used to
determine \cite{DN} $K$ matrices for arbitrary spin.  In particular,
the coefficients $c_{p \,, l}$ appearing in Eq.  (\ref{nfunction}) are
given by
\be
c_{p \,, l} = {(p+1)\over l!} \prod_{k=0}^{l-2} (p-l-k) \,. \non 
\ee
Thus, for $p \rightarrow \infty$, $c_{p \,, l} \sim {p^l\over l!}$.

\end{document}